# Deep *Chandra* Observations of ESO 428-G014: IV. The Morphology of the Nuclear Region in the Hard Continuum and Fe Kα Line

G. Fabbiano[a], A. Siemiginowska[a], A. Paggi[a, b,c,d], M. Elvis[a], M. Volonteri[e], L. Mayer[f], M. Karovska[a], W. P. Maksym[a], G. Risaliti[g,h], Junfeng Wang[i]

a. Center for Astrophysics, Harvard & Smithsonian, 60 Garden St. Cambridge MA 02138, USA

b. Dipartimento di Fisica, Universita' degli Studi di Torino, via Pietro Giuria 1, I-10125 Torino, Italy

c. Istituto Nazionale di Fisica Nucleare, Sezione di Torino, via Pietro Giuria 1, 10125 Torino, Italy

d. INAF-Osservatorio Astrofisico di Torino, via Osservatorio 20, 10025 Pino Torinese, Italy

e. Institut d'Astrophysique de Paris, Sorbonne Universit´es, UPMC Univ. Paris 06 et CNRS, UMR 7095, 98bis boulevard Arago, F-75014 Paris, France

f. Center for Theoretical Astrophysics and Cosmology, Institute for Computational Science, University of Zurich, Winterthurerstrasse 190, CH-8057 Zu¨rich, Switzerland

g. Dipartimento di Fisica e Astronomia, Università di Firenze, via G. Sansone 1, I-50019 Sesto Fiorentino (Firenze), Italy

h. INAF - Osservatorio Astrofisico di Arcetri, Largo E. Fermi 5, I-50125 Firenze, Italy

i. Department of Astronomy and Institute of Theoretical Physics and Astrophysics, Xiamen University, Xiamen, 361005, China

Abstract

We report the results of high-resolution subpixel imaging of the hard continuum and Fe Kα line of the Compton Thick (CT) Active Galactic Nucleus (AGN) ESO 428-G014, observed with *Chandra* ACIS. While the 3-4 keV emission is dominated by an extended component, a single nuclear point source is prominent in the 4-6 keV range. Instead, two peaks of similar intensity, separated by ∼36 pc in projection on the plane of the sky are detected in the Fe Kα emission. The SE knot could be marginally associated with the heavily obscured hard continuum source. We discuss four possible interpretations of the nuclear morphology. (1) Given the bolometric luminosity and likely black hole (BH) mass of ESO 428-G014, we may be imaging two clumps of the CT obscuring torus in the Fe Kα line. (2) The Fe Kα knots may be connected with the fluorescent emission from the dusty bicone, or (3) with the light echo of a nuclear outburst. (4) **We also explore the less likely** possibility that we may be detecting the rare signature of merging nuclei. Considering the large-scale kpc-size extent of the hard continuum and Fe Kα emission (Papers I and II), we conclude that the AGN in ESO 428-G014 **has been active for at least $10^4$ yrs. Comparison with the models of Czerny et al (2009) suggests high accretion rates during this activity.**

# 1. Introduction

Compton Thick (CT) active galactic nuclei (AGN) have been considered as a possible crucial stage in the joint supermassive black hole (SMBH) – galaxy evolution. The M-σ relation (Magorrian et al. 1998) suggests that the evolution of galaxies and their nuclear supermassive black holes (SMBHs) are linked. Both the stellar population and the SMBH are thought to grow and evolve by merging of smaller gas-rich galaxies and their nuclear SMBHs (Daddi et al. 2007). During this process, the SMBHs are likely to merge while 'buried' by thick molecular gas, which accretes at high rates, causing the birth of an – initially obscured - AGN. The awakened AGN, in turn, may exercise feedback on the galaxy by means of radiation and winds, blowing away the surrounding gas and quenching star formation (Di Matteo, Springer and Hernquist 2005).

*Chandra* observations of nearby CT AGNs are beginning to resolve the nuclear region, showing unexpected complexity. In the nearby CT AGN NGC 4549 (Marinucci et al. 2012, 2017) and the Circinus galaxy (Arevalo et al. 2014) flattened ~200pc circumnuclear distribution of CT scattering and fluorescing clouds have been reported, suggesting an association with the obscuring torus of the 'standard' AGN model. Other CT AGNs have been found in double-nucleus merging galaxies (e.g., first in NGC 6240 Komossa et al. 2003), consistent with the idea that CT AGN may be a manifestation of the merging cycle of galaxy evolution.

Our deep *Chandra* data set on ESO428-G014 provides a unique opportunity for an exploration of the circumnuclear region of a CT AGNs. ESO428-G014 (also called IRAS01745-2914, MCG-05-18-002), is a southern barred early-type spiral galaxy [SAB(r)], at a distance of ~23.3 Mpc (NED; scale=112 pc/arcsec). This galaxy has a highly obscured CT ($N_H > 10^{25}$ cm$^{-2}$) Seyfert Type 2 nucleus, with a high ratio of [OIII] λ5007 to hard 2-10 keV observed flux (Maiolino et al. 1998), and the second highest [OIII] flux among CT AGNs after NGC1068 (Maiolino & Rieke 1995, Risaliti et al. 1999), so it is expected to be the second intrinsically brightest CT AGN below 10 keV.

ESO 428-G014 is comparable to NGC 4151 in both black hole mass and accretion rate. The likely black hole mass[1] in ESO 428-G014 is $(1 - 3) \times 10^7$ M$_\odot$, using the McConnell and Ma (2013), scaling from bulge velocity dispersion, σ, and mass, M(K)$_{bulge}$, with the Peng et al. (2006), values of σ = 119.7 km s$^{-1}$ and M(K)$_{bulge}$ = $10^{9.14}$ M$_\odot$, and assuming a mass to light ratio of 0.6, (McGaugh and Schombert 2014). This is a similar mass to that of NGC 4151 (4.5 x 10$^7$ M$_\odot$, Bentz et al., 2006). Given this black hole mass estimate, the Eddington ratio of ESO 428-G014 is about 1%. This is comparable to the 1.2% value for NGC 4151 (Wang et al., 2011).

In our previous work on this AGN we have reported: (1) the discovery of kiloparsec-scale extended components in the 3-6 keV hard continuum and Fe Kα 6.4 keV line emission, suggesting scattering in the interstellar medium (ISM) of photons escaping the nuclear region (Paper I, Fabbiano et al. 2017); (2) a detailed spectral and spatial study of the diffuse large-scale kpc-size emission in the entire *Chandra* energy range (Paper II, Fabbiano et al. 2018a); and (3) a detailed study of the inner ~500 pc radius high-surface brightness emission, in comparison with optical line and radio properties (Paper III, Fabbiano et al. 2018b).

---

[1] ESO 428-G014 does have a VLBI water maser (Zhang et al., 2006), but does not have a maser-derived black hole mass.

In this fourth and final paper we report the results of a high angular resolution *Chandra* study of the inner (~100 pc diameter) nuclear region of ESO428-G014 at the energies > 3.0 keV, where the hard featureless continuum and Fe Kα line emission are prevalent.

## 2. Data Reduction & Analysis

Table 1 summarizes the observations used in this paper. The data and basic reduction and analysis procedures used here follow closely those used in Papers I, II and III. We refer to those papers for detailed descriptions of some aspects of the analysis. We repeat here the basic procedures, and add specific information pertinent to this paper.

Table 1. Observation Log

| ObsID | Instrument | $T_{exp}$(ks) | PI | Date |
|---|---|---|---|---|
| 4866 | ACIS-S | 29.78 | Levenson | 2003-12-26 |
| 17030 | ACIS-S | 43.58 | Fabbiano | 2016-01-13 |
| 18745 | ACIS-S | 81.16 | Fabbiano | 2016-01-23 |

As explained in Paper I, all the data sets were screened, processed to enable sub-pixel analysis and merged. For our analysis, we used *CIAO* 4.8 tools (http://cxc.cfa.harvard.edu/ciao/) and the display application *DS9* (http://ds9.si.edu/site/Home.html), which is also packaged with *CIAO*. To highlight some of the features, we applied different Gaussian kernel smoothing to the images, as described in the figure captions.

The *Chandra* PSF was simulated using rays produced by the *Chandra Ray Tracer* (*ChaRT*) (http://cxc.harvard.edu/ciao/PSFs/chart2/) projected on the image plane by *MARX* (http://space.mit.edu/CXC/MARX/). We have also applied the *LIRA* software package (github.com/astrostat/LIRA, Van Dyk et al. 2016; see also McKeough et al. 2016), to estimate the statistical significance of a nuclear double peak in the 6-7 keV range (Section 3.2). As in Papers II and III, we have applied the image restoration algorithm Expectation through Markov Chain Monte Carlo (*EMC2*, Esch et al. 2004; Karovska et al. 2005, 2007; sometimes referred to as PSF deconvolution) with the appropriate PSF model (Section 4).

In Paper I we discussed the method followed for merging the three observations (Table 1). We summarize it here as well, to reinforce the point that astrometry problems cannot cause the 6-7 keV double source we will discuss in Sections 3 and 5. We first inspected visually each image in the 0.3–8.0 keV band and established that possible centroid shifts were below one ACIS instrument pixel (0".492). We then used detected X-ray sources in each field to cross-match the images, excluding the central source of ESO 428-G014. Finally we inspected the images again and made corrections (< ½ ACIS pixel in the worst case), to ensure that the nuclear centroids were appropriately shifted, considering both the hard continuum and the Fe K band. Fig. 1 shows the nuclear source in the 4-6 keV range (left two columns) and 6-7 keV range (right two columns). In each case we image each observation (prior to the shifts; identified by the obsid in the upper left of each panel) and the merged image. We plot with a circle the position of the 4-6 keV nuclear source in the longest observation (i.e. best statistics, obsid 18745), as reference point. Obsid 18745 shows a single well-defined peak in the 4-6 keV band, and two separate peaks in the 6-7 keV band. The shorter observations also show single peaks in the 4-6 keV band, and within statistical variations, two peaks or elongations in the 6-7 keV band. Similarly, the merged images show a single peak in the 4-6 keV band and a double peak in the 6-7 keV band.

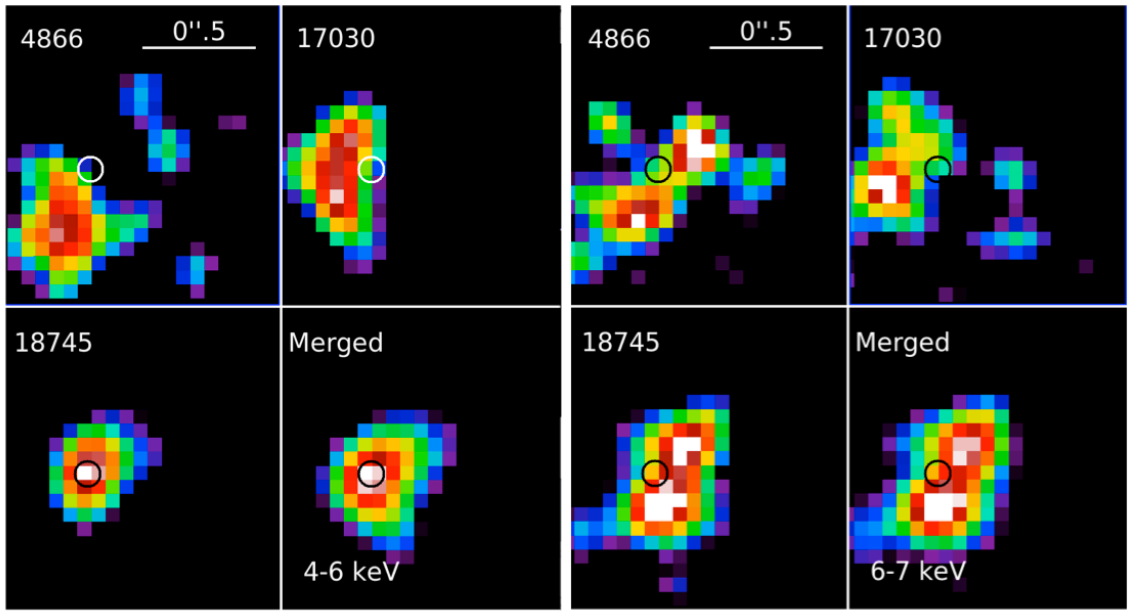

Fig. 1 - The nuclear region in individual and merged observations. The four panels on the left show the 4-6 keV band prior to shifting, and the merged image; the four panels on the right show the 6-7 keV band. The circle corresponds to the centroid of the 4-6 keV source in the longest single observation (18745). The color scale was stretched to emphasize the centroids; white are the highest count pixels. The data was binned in 1/8 of the 0".492 ACIS instrumental pixel. A Gaussian smoothing with 3 image pixel kernel was applied.

## 3. Morphology of the Hard (> 3 keV) Nuclear Emission: the 'Point-like' Unresolved Component

The spectral analysis of Paper II has shown that the emission between 3 and 6 keV is a featureless hard continuum. A strong neutral Fe Kα line with the contribution of a much weaker Fe XXV line dominates the spectrum in the 6-7 keV band. **In NGC 4945, using narrow spectral bands, Marinucci et al (2017) studied the spatial distributions of the neutral and ionized Fe Kα emission separately. We have explored energy dependent behaviors in ESO 428-G014 by imaging separately the counts in the 6.2-6.5 keV band (neutral line) and 6.5-6.9 keV band (ionized line), but we find similar double-clump morphologies in both bands. Hence, to improve statistics, we consider the entire 6.0-7.0 keV band, unless otherwise specified.**

Fig. 2 shows images of the central ~2'' region in four energy bands above 3 keV. For these images the data were binned at sub-pixel resolution of 1/8 and then smoothed with a Gaussian kernel of 2 image pixels.

The positions of the peak of the nuclear emission are spatially coincident in the 4-5 keV and 5-6 keV bands (top-right and bottom-left in Fig. 2). The centroid of the source does not change up to 6.2 keV. This position (J2000 RA = 7:16:31.21, Dec = -29:19:28.6), which we assume as the true nuclear position, is consistent with that used in Paper II as the center of the spectral extraction regions. In the standard model of a CT AGN, these photons are nuclear photons scattered and reflected by the obscuring circumnuclear clouds. The morphology of the peak emission is consistent with a single nuclear point source.

In the 3-4 keV band (Fig. 2, top-left panel) the centroid is less prominent and appears to migrate spatially to the SE, suggesting that the nucleus is obscured at these energies and other phenomena may be at play, either connected with the radio jet or line emission in the ISM (see Paper III).

The 6.0-7.0 keV band (Fig. 2, bottom-right panel) contains both neutral and ionized Fe Kα lines, as well as continuum. Paper II shows that the emission is dominated by the large Equivalent Width (EW) neutral Fe Kα line at 6.4 keV (EW 1.6±0.5 keV, Levenson et al 2006). Compared with the single-peaked 4-6 keV continuum source, the 6-7 keV emission appears elongated or double. The "double" source starts to be visible at energies > 6.2 keV, suggesting that it is due to the Fe Kα line. The two peaks in the Fe Kα image are of similar intensity, with 44 and 38 counts in 0''.16 radius extraction circles for the southern and northern peak respectively. The centroids are separated by ~0''.32.

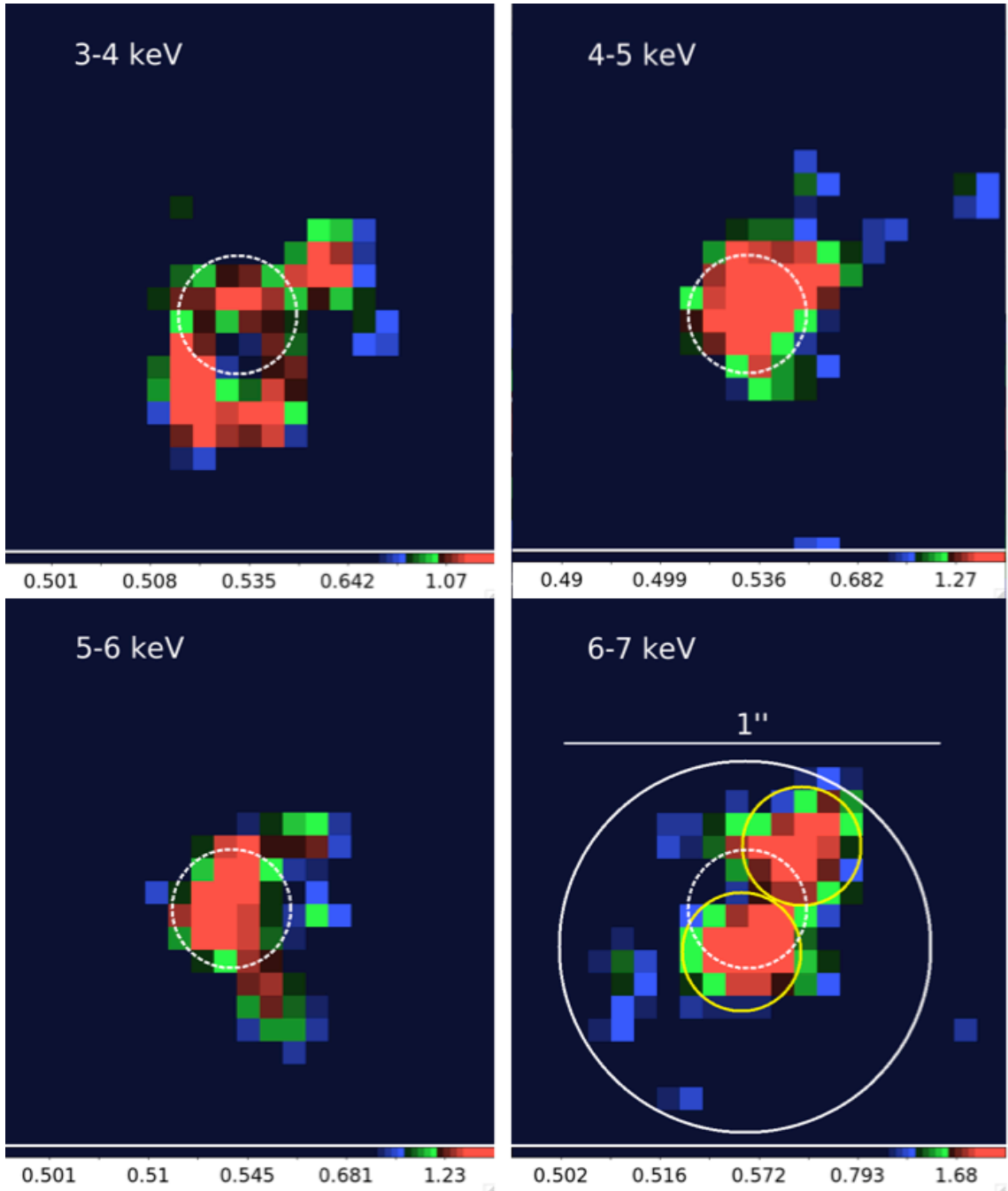

Fig. 2 –Images of the nuclear region in the indicated energy bands, in the same spatial scale. The data were imaged with 1/8 of the ACIS pixel resolution and smoothed with a 2 image-pixel Gaussian. N is to the top and E to the left. The dashed white circle marks the centroid of the 4-6 keV source. The yellow circles are the 0".16 radius circles used for count extraction in the 6.0-7.0 keV band. The white circle has 0".5 radius. The image intensity scale is logarithmic. The color scale is in counts per 1/8 ACIS pixel.

The centroid of the southern 6-7 keV peak is displaced from that of the 4-6 keV continuum by 0".12. Given the counting statistics in each peak and the *Chandra* PSF FWHM, we estimate that the error on this is 0".03, so formally the displacement is at a 4σ significance. If we take this at face value, the southern peak of the Fe Kα emission may not be coincident with the peak of the nuclear continuum source.

Given that the 4-6 keV nuclear source is single, the double 6-7 keV source cannot be the results of bad astrometry in the merging of the individual observations of ESO 428-G014 (See also Section 2).

### 3.1 Could PSF systematics cause the double Fe Kα emission?

The position and relative intensity of the northern source exclude that this double peak may be due to the well-documented PSF artifact [2]. However, a 2004 Calibration memo based on an observed ACIS-S HETG zeroth-order source suggests that the hard PSF (6 - 9 keV) could have a double appearance in the central ~0".5 (Jerius et al. 2004), with a peak separation of ~0".23 (so contained within one ACIS instrumental pixel). In the 6-7 keV band, the mirror PSF (derived with *ChaRT* from pre-flight calibrations) does not show any deviation from a single-peak response. Instead it shows significant departure from a single peak above 7 keV, which is above the energies under consideration here. Both bands are included in the Jerius et al 6-9 keV in-flight calibration. Fig. 3 shows the mirror PSF we have simulated for the 6-7 keV band, with 1/8 pixel binning. Superimposed are two circles corresponding to the yellow circles in Fig. 2. The 6-7 keV PSF is clearly single peaked.

To investigate potentially unknown 'flight systematics' of the 6-7 keV PSF, that may cause a double peaked source as we observe in the 6-7 keV band, we used the archival ACIS-S observations of the AGN NGC 4507 (ObsID 12292, no gratings), which has a prominent source at the aim-point. Fig. 4 shows these data plotted in the 4-6 and 6-7 keV bands. Within a 1" radius circle this source yields ~2470 counts in the 4-6 keV band and ~1480 counts in the 6-7 keV band, so that statistics are excellent. The observation was taken in ¼ subarray mode, so that pileup is not an issue. The data were analyzed as for Fig. 2. The white circle has a 0".16 radius as in Fig. 2 and is centered on the 4-6 keV centroid. In this case a single source with the same centroid is visible in the 6-7 keV band. We thus conclude the observed ACIS-S PSF at the aim-point does not produce a double peak in the 6-7 keV band. This gives us confidence in using the single-peaked *ChaRT* 6-7 keV model PSF as a template for our simulations (see below).

We further checked the NGC 4507 image in the 7-8 keV band. At these energies, indeed a secondary peak appears, as suggested by the mirror PSF, and may explain the in-flight calibration results for the 6-9 keV energy band.

---

[2] http://cxc.harvard.edu/ciao/caveats/psf_artifact.html

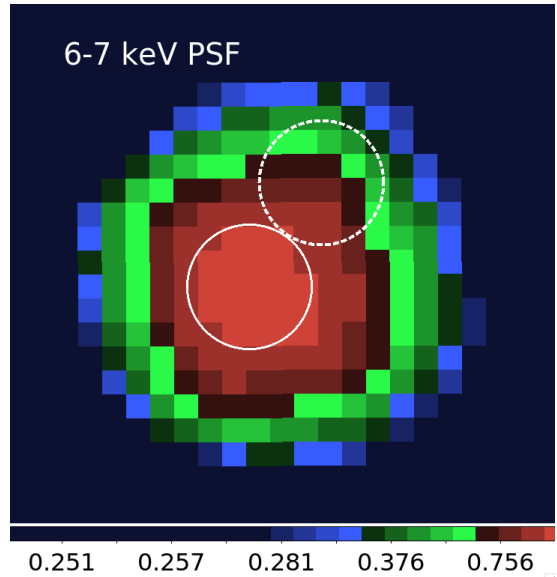

Fig. 3 – Mirror PSFs for the 6-7 keV energy band. The PSF is binned in 1/8 of ACIS pixel, smoothed with a 2pix Gaussian and displayed in a logarithmic intensity scale, as we do for the data. The circles have radius of 0".16.

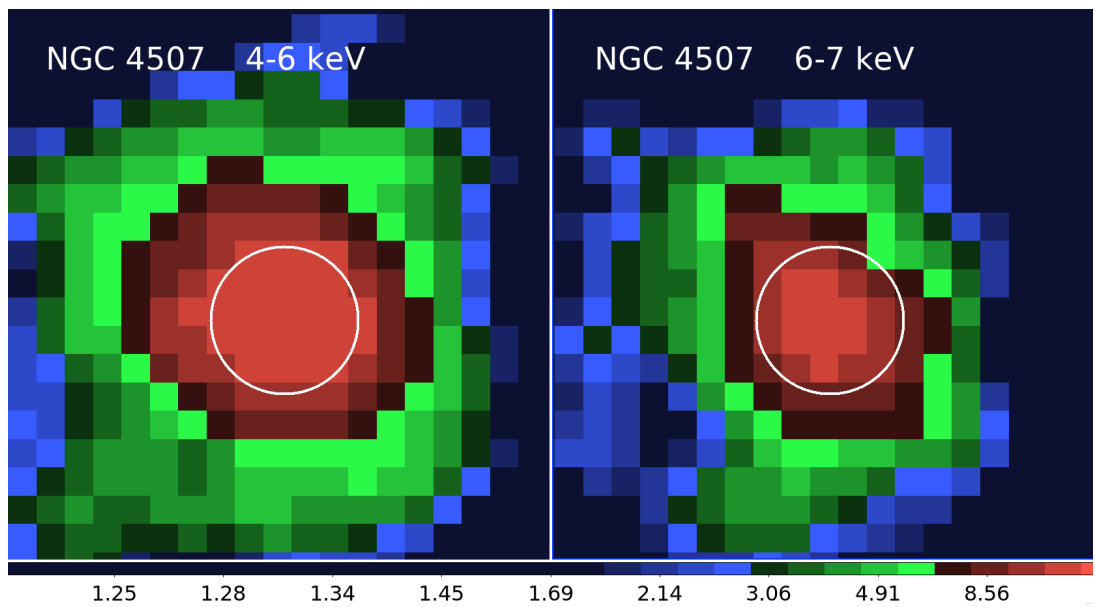

Fig. 4 - Images of the nuclear source of NGC 4507 in the indicated energy bands, in the same spatial scale. N is to the top and E to the left. The white circle (0".16 radius) marks the centroid of the 4-6 keV source. The centroid of the PSF in the 6-7 keV band is consistent with that of the 4-6 keV band, and there is no second source, thus excluding a PSF 'feature' for the double source in ESO 428-G014. The image intensity scale is logarithmic. The color scale was stretched in each image so to highlight the inner core. The data were imaged with 1/8 of the ACIS pixel resolution and smoothed with a 2 image pixel Gaussian.

## 3.2 Statistical significance of the double Fe Kα feature

We have approached the evaluation of the statistical significance of the double feature in two ways. First, directly from the image, taking counting statistics into account and using the 6.4 keV *Chandra* calibration encircled energy curves. Second, by applying the statistical software package *LIRA* (Stein et al 2015). *LIRA* is based on the same approach we use to generate the *EMC2* reconstructed images in Paper III (see also below). This method uses the full *Chandra* PSF and will take into account small-scale asymmetries that would be averaged over in the encircled energy approach. In the latter case a PSF model was generated for the 6-7 keV energy band and used as statistical null. A note of caution is that in both approaches the results are based on the same available PSF calibrations, and both will be similarly affected by unknown systematics in the knowledge of the PSF core. However, as discussed above (Section 3.1, see Fig. 3) there is no evidence of a double peaked response, such as we observe, originating from such unknown systematics.

### 3.2.1 Direct estimation from the image

For the direct approach, we extracted counts from two separated circles of radius 0''.16 centered on the two 'sources' visible in the image. We take as our reference the southern source, which is positionally closer to the 4-6 keV continuum source. Using the 6.4 keV PSF encircled energy curve, we then calculated the expected counts from this PSF in a circle of 0''.5 radius, which includes the northern source (see Fig. 2, lower-right). We predict 143 counts from the southern source in this circle, and we detect 203 counts. The difference is easily reconciled, within statistics, to the presence of the northern source, using the same encircled energy approach.

We then asked the question, given the southern source as true, what is the significance of the northern source? We estimate that the background from the PSF spread counts of the southern source is ~4.5 counts in a 30 degrees annular sector containing the northern source. Given the 38 counts we derive from the image for the northern source, the net source counts would be 33.5±6.5, or 5.1 σ significance.

Alternatively, given the possibility of local diffuse emission (Section 4), we also estimated the background locally, from a circle of 0''.16 radius SW of the northern peak. The local estimate of the background is 12 counts; given the presence of the two points sources, most of these counts (if not all, within statistics), would be due to counts from these sources spread by the PSF. In this overly conservative approach, we obtain 26±7 counts for the northern source, or a 3.7σ significance.

### 3.2.2 Upper limit on the chance probability from *LIRA*

The second approach to test the statistical significance of a double nuclear Fe Kα source in ESO428-G014 was to apply the method presented in Stein et al. (2015). We calculated the upper bound on the p-value using the results from applying the *Low-counts Image Reconstruction and Analysis* (*LIRA*) software package ([github.com/astrostat/LIRA](github.com/astrostat/LIRA), Van Dyk et al. 2016; see also McKeough et al.

2016). *LIRA* has been developed to study sub-PSF structures using a multi-scale spatial model, accounting for the PSF, allowing for multiple image components (e.g., background, known point sources, unknown extended structure), and nested within a fully Bayesian framework (Esch et al. 2004, Connors and Van Dyk 2007, Connors et al. 2011). *LIRA* is not a deconvolution algorithm, but a forward-fitting process that is robust to counts fluctuations and numerical instability. *LIRA* results contain a series of Markov chain Monte Carlo (MCMC) images drawn from the posterior distribution of the model given the observed and simulated data. These images are then used in the calculation of the test statistics and the upper bound on the probability for any structures present in the assumed a priori regions.

Using this approach, we investigated the statistical significance of a departure from a baseline single point source model in the 6-7 keV band, given the Poisson noise of our data, the 6-7 keV *Chandra* PSF, and the detector inefficiencies. The test statistic of this method is based on the tail probability of the posterior distribution under the test model, and yields an upper bound on the probability under the null hypothesis of observing in a simulated image a secondary source at the same level as in the observed image (see Eq. 10 in Stein et al. 2015). This can be interpreted as the probability that a model comprising a single point source would yield the observed image by Poisson fluctuations.

We simulated 100 images (128×128 pixels, 1pix = 0".0615) using *Sherpa* (Freeman et al. 2001), assuming a point source centered on the nucleus [(RA, Dec) = (07:16:31.209, -29:19:28.73)] plus a Poisson background, and a *ChaRT* simulated PSF in the 6-7 keV energy range. We then used *LIRA* to obtain 2000 MCMC images from the posterior distribution of each null model image and the observed *Chandra* image and calculated the parameter $\xi$ defined in Stein et al. (2015), which characterizes the proportion of the total intensity due to emission not present in the null model. Figure 5 shows the posterior distributions of the parameter $\xi$ in the secondary source region [centered on (RA,Dec)= (7:16:31.195, 29:19:28.43)] for the observed image and the simulated images under the null model. The distribution of the $\xi$ parameter for the data (blue curve in Fig. 5) indicates a high proportion of the counts in this region in comparison to the null images (gray and black).

We then calculated the upper limit to the probability that the presence of a secondary source in the nuclear region of ESO428-G014 may be due to statistical fluctuations. We obtain P<0.0016. We conclude that the difference between the 4-6 keV and the 6-7 keV nuclear emission is real at high confidence.

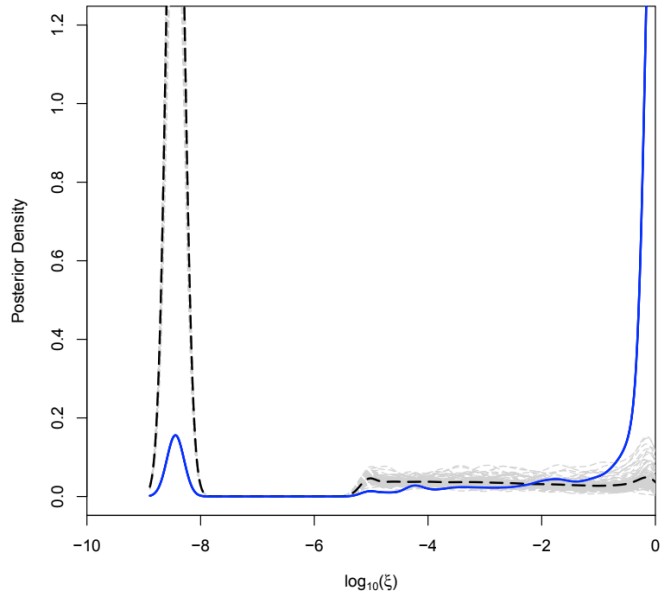

Fig. 5 – The posterior distributions of the parameter ξ (Stein et al 2015, see text) in the north region [centered on (RA,Dec)= (7:16:31.195, 29:19:28.43)] for the observed image (blue solid curve), and the 100 (light grey curves) simulated images under the null model, which includes only one nuclear source (coincident with the southern point source) and a background. The dashed black curve is the average across all the nulls. Note the difference between the large peaks in the blue (data) and black (nucleus only model) curves.

## 4. The Extended Environment of the Double Fe Kα peaks

The 6-7 keV emission over a slightly larger (~3'', 338 pc) region is shown in Fig. 6. Fig. 6 suggests that the two nuclear Fe Kα peaks are embedded in a wider, and possibly structured low-surface brightness Fe Kα emission, the central component of the large-scale Fe Kα emission discussed in Papers I and II. As in Paper III, we used 1/16 pixel sub-binning, to highlight the structure of the diffuse emission. The right panel of Fig. 6 shows the 6-7 keV raw image, adaptively smoothed with Gaussian kernels ranging from 0.5 to 1.5 pixels. While this image shows counts spread over a ~1'' radius, a good fraction of these counts are expected from the PSF spreading the counts of the two peaks sources (see Section 3.2.1). The left panel shows the same data after *EMC2* image reconstruction, (which corrects for the effect of the *Chandra* PSF), similarly smoothed. The circles indicate the positions of the nuclear peaks in the raw data (see Fig. 2). The *EMC2* reconstructed image shows the beginning of the large scale elongated emission along the ionization cone, discussed in our previous papers, and a central rounder region with a radius of ~1" ~ 112 pc. Within this area resides a central brighter and apparently structured region of ~50pc radius, which includes the two nuclear peaks discussed in Section 3.

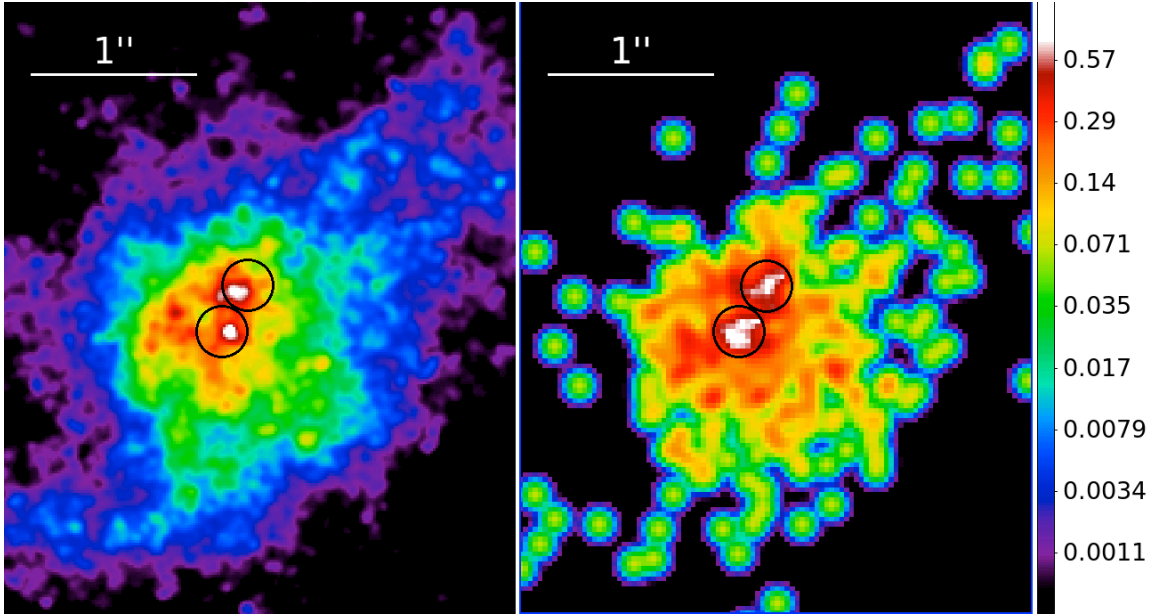

Fig. 6 – 6-7 keV, 1/16 ACIS pixel images of the nuclear region. Left: *EMC2* reconstructed image. Right: raw image (the single counts along the diagonal belong to the lower surface brightness very extended Fe Kα emission discussed in Papers I and II). Both images are slightly smoothed (see text). The color-intensity scale to the right refers to the raw data. The back circles highlight the two nuclear peaks discussed in this paper.

5. Discussion

We have found that the morphology of the nuclear region changes within the 3-7 keV spectral range. Spectrally, this band is dominated by the continuum emission and the Fe Kα neutral line, with the Fe XXV line contributing ~20% of the neutral line flux (Paper II).

Fig. 2 shows that a nuclear continuum point source becomes prominent only at energies > 4 keV. As mentioned in Section 3, this single peak can be seen up to ~6.2 keV where the neutral Fe Kα line begins to dominate the emission. In the 6-7 keV energy band the emission appears split into two knots, separated by ~36 pc in projection. The southern knot is near to the 4-6 keV continuum source, although the centroids differ by 13 pc in projection, at 4σ significance. Both knots reside in a region of lower surface brightness extended Fe Kα emission (Fig. 6). Below we examine the implications of our results for four possible interpretations of the nuclear morphology (Sections 5.1 – 5.4). We then discuss constraints imposed by the very extended kpc-size emission (Paper II) for the lifecycle of the AGN (Section 5.5).

**5.1 Imaging the AGN torus?**

Both the >3 keV continuum and the narrow, large EW 6.4 keV Fe Kα emission of CT AGNs are believed to originate from the scattering of hard photons from near the central black hole with a ~1-100 pc scale dense circumnuclear 'torus' (e.g. Gandhi et al. 2015).

The best-studied small-scale nuclear morphology in a CT AGN is that of the nearby galaxy NGC 4945, where ~200 × 100 pc flattened distributions of hard continuum, and both neutral and ionized Fe Kα lines have been discovered with *Chandra*, perpendicular to the ionization cone of this AGN (Marinucci et al. 2012, 2017). Marinucci et al. (2012) ascribed this emission to the AGN obscuring torus. A similar flattened distribution (~160 pc across) is found in the Circinus galaxy (Arevalo et al. 2014). The Circinus feature has a spectrum including soft line dominated emission, hard continuum and Fe Kα line, but imaging in different energy bands is not available in the literature.

In ESO 428-G014, instead, we do not detect a flattened feature, although there may be a suggestion in the ratios of hard and soft band images (Paper III). What we have is a structured roughly round low surface brightness region, surrounding two bright, unresolved emission peaks (Fig. 6). However, some similarity with NGC 4945 exists. In particular Marinucci et al. (2017) observe that on smaller scales within the ~200 pc feature of NGC 4945, localized excesses of continuum, neutral and ionized Fe Kα emission are not correlated. These authors suggest that this is the result of the complex clumpy cloud distribution surrounding the AGN. The same may occur in ESO 428-G014. In NGC4945 there is strong Fe Kα emission from within ~10 pc of the nucleus, possibly in contrast to ESO 428-G014 (Section 3). Marinucci et al. (2012) note that a cylinder seen nearly edge-on will produce a bright central spot, without any emission from the nucleus, so the two objects could be intrinsically similar, and the observed difference just a matter of inclination with respect to our line of sight.

It seems likely, however, that the extended hard X-ray region found by Marinucci et al. (2012) in NGC 4945 is not the standard obscuring torus of AGN models. The X-ray source luminosity in NGC 4945 is intrinsically L(2-10 keV) ~ $10^{42}$ erg s$^{-1}$ (Iwasawa et al., 1993, Koss et al., 2017, scaled to a 2-10 keV luminosity using PIMMS[3] and Γ = 1.8), ~1/50 as luminous as in ESO 428-G014. The inner edge of the obscuring torus will lie at $r_{1500}$ ~ 0.014 pc, where the dust equilibrium temperatures is ~1500 K (where r = 1.3 $L_{UV,46}^{0.5}$ $T_{1500}^{-2.8}$ pc, Barvainis 1987). The H$_2$O maser implies a central black hole mass of ~1x10$^6$ M$_\odot$ (Greenhill et al., 1997), which puts the inner edge at ~10$^5$ Schwarzschild radii. The IR AGN emission reaches to ~20 microns (Asmus et al. 2015), so down to dust temperatures of ~150 K. Given the steep dependence of radius on dust temperature, the radius $r_{150}$, where the torus dust reaches 150 K, would then be $r_{150}$ = 600 $r_{1500}$, i.e. ~8 pc, much smaller than the ~100 pc radius region seen in hard continuum and Fe Kα by Marinucci et al. (2012) in NGC 4945. The above considerations suggest that the Marinucci et al. structure may be part of the interstellar medium of the host galaxy. As discussed in Papers I, II and III, molecular clouds in this medium may quite efficiently scatter hard nuclear photons to produce the observed kpc-scale hard continuum and Fe Kα emission.

---
[3] https://heasarc.gsfc.nasa.gov/cgi-bin/Tools/w3pimms/w3pimms.pl

In contrast, the two observed knots of Fe Kα emission in the central 100 pc of ESO 428-G014 are closer to the expected radii for the torus than in NGC 4945. At the intrinsic AGN bolometric luminosity of ESO 428-G014 (4.1×10$^{43}$ erg s$^{-1}$, Levenson et al., 2006), the inner edge of the obscuring torus is $r_{1500}$ ~ 0.1 pc, and the outer radius, $r_{150}$ ~50 pc. The compact knots (<18 pc radius) of ESO 428-G014 have similar luminosity. Neither is obviously coincident with the heavily obscured continuum source seen at 4 - 6 keV. The Southern Fe Kα region is offset by ~13 pc at ~4 σ significance, the Northern Fe Kα region is offset by ~36 pc at high significance. **Given the statistics of our data, if the Southern offset is real, we estimated that at a 3σ upper limit the Fe Kα from the obscured continuum source may be only ~20% of the total in this region.**

However, the Northern Fe Kα knot is only ~30 degrees from the NLR "bicone" axis. This angle is not expected for the torus, which should be at ~90° to the bicone axis.

## 5.2 Features of the biconical outflow?

The Northern Fe Kα knot is at an ambiguous angle of ~30° to the bicone axis. At this angle it may lie within or outside the bicone. If outside, then this region may simply be fluorescing because it is illuminated by the central X-ray source through a gap in the clouds making up the torus (Nenkova et al. 2008). If instead the northern Fe Kα knot lies inside the bicone, then it will certainly be illuminated by the central X-ray source, but is not then a part of the obscuring torus. The question is then whether the fluorescing cloud is simply another ISM molecular cloud or is related to the biconical outflow.

We know that there is dust within the bicone. The mid-IR (12 micron) emission from ESO 428-G014 is extended along the same position angle as the bicone (Asmus et al. 2016), rather than perpendicular to the bicone, as expected for torus dust. Moreover, the presence of large-scale diffuse hard continuum and Fe Kα emission argues for a molecular cloud, and so dusty, scattering medium in the bicone (Papers I and II). A clear example of a fluorescing bicone can be found in the Circinus galaxy (Arevalo et al. 2014).

## 5.3 The echo of a recent past nuclear outburst?

In the Milky Way, localized reflection features within 100 pc from Sgr A* have been interpreted as the X-ray echo from a nuclear outburst interacting with molecular clouds (Koyama et al., 1996, Ponti et al., 2015). Churazov et al. (2017b, c) present a time evolution model of the X-ray morphology assuming a factor of 10 outburst (from $L_X=10^{40}$ erg s$^{-1}$) lasting for 50 or 5 years. The two illuminated regions would be visible within the first 200 years, before becoming diffuse and disappearing. The timescale for the evolution of the illuminated region depends on the distribution of the clouds and their densities. More luminous outbursts can produce more compact regions initially. It is possible we are seeing such an effect in ESO 428-G014. Given the ~30 pc projected separation of the Fe K$\alpha$ knots, the light travel time is >~90 years.

### 5.4 A double nucleus?

CT AGNs have the characteristics predicted for the 'buried' pre-blow-out phase in the merger cycle (Hopkins et al., 2008). An increasing number of dual CT AGNs have been reported from *Chandra* observations of actively merging galaxies. Starting with the two Fe-K$\alpha$ sources with ~690 pc separation in NGC 6240 (Komossa et al. 2003), evidence for double CT AGNs has been reported in Mkn 463 (~3.8 kpc separation, Bianchi et al. 2008), and possibly in Arp 299 (~4.6 kpc separation, Ballo et al. 2004), and Arp 220 (~400 pc separation, Paggi et al. 2017).

In ESO 428-G014, we have a post-merger CT AGN, with two clearly identifiable at high statistical significance neutral Fe K$\alpha$ sources in the innermost region, separated by ~30 pc. As discussed above, these sources may be explained in the context of reflection and scattering of nuclear photons from two different CT clouds, in the AGN torus itself or in the host galaxy ISM. **Here, for completeness, we also discuss the possibility of there being two, still-merging, SMBHs.**

**Although unlikely, we cannot exclude that the SE source could be the counterpart of the hard continuum nucleus, given the -statistical only- 4$\sigma$ confidence on the displacement (Section 3). Using a 6.2-6.5 keV band to extract the Fe K$\alpha$ neutral line counts, and the same extraction circle as in Section 3, we obtain 20 net counts in the NW Fe K$\alpha$ clump and a 3$\sigma$ upper limit of 9.5 counts for a 5.0-6.0 keV source, so an Fe K$\alpha$ EW of >2 larger. Instead, for the SE source, assuming that it is coincident with the continuum peak, we obtain EW~0.9 keV.** If the NW source were a second AGN, given the count statistics, it may have a factor of >2 larger EW than the source corresponding to the hard continuum peak, suggesting a more heavily cocooned CT source.

Simulations of mergers, including both the stellar and the gaseous components (e.g., Callegari et al. 2011) predict accretion onto the SMBHs of the parent galaxies, resulting in double CT AGNs. Since the majority of low-z galaxies are spirals, merging will most likely involve two spiral galaxies. Using isolated merger simulations of spiral galaxies, Capelo et al. (2017) study the level of obscuration caused by the ISM on scales >20-40 pc and the observability of dual AGN. In these simulations, the primary galaxy has a stellar mass of ~$1 \times 10^{10}$ M$_\odot$ and the secondary a mass between $1 \times 10^9$ and $1 \times 10^{10}$ M$_\odot$, covering mass ratios in the range 1:10 to 1:1. The column density of gas they find at scales >30 pc during the merger is between $1 \times 10^{23}$ and a few times $1 \times 10^{24}$ cm$^{-2}$; additional obscuration by clouds/torus near the black hole would increase these values (see Paper II and Section 5.1).

The typical time over which a second AGN, with a luminosity as low as 1-10% of the main AGN (assuming for the latter ~4×10$^{43}$ erg s$^{-1}$ as in ESO 428-G014), could be observed within 100 pc is between 0.1 and 5 Myr depending on the merger mass ratio and orbit, for SMBHs with mass ~1×10$^6$ M$_\odot$. By scaling the duration with the Eddington ratio (see supplementary figures in Capelo et al. 2017) this timescale would be longer (shorter) for more (less) massive BHs, up to more than 1×10$^7$ years for BHs with mass ~1×10$^8$ M$_\odot$. Given our estimate in Section 1 for the BH mass, $(1 – 3) \times 10^7$ M$_\odot$, the timescale may be somewhat in between 5-10 Myr, so that although unlikely, we cannot completely rule out the dual BH hypothesis.

**5.5 The nucleus and the large-scale extended emission**

**In Paper I of this series, we have reported that the hard continuum and Fe K$\alpha$ emission extend out to 3-5 kpc from the nucleus and we suggested that this emission is due to the interaction of hard nuclear photons with dense molecular clouds in the disk of the host galaxy. In Paper II we analyzed the spectrum of the extended emission from 0.3 to 8.0 keV and found best fits with composite photoionization + thermal models. In Paper III, this spectral complexity is partly reconciled with morphological features, where we study in detail the morphology of the diffuse X-ray emission in the inner ~500 pc radius region, finding localized differences in the absorbing column and in the emission processes. In particular, we find that collisional ionization may be prevalent in the area of most intense optical line emission (H$\alpha$ and [OIII]), which corresponds to areas of interaction with the nuclear radio jet.**

**In Paper II we also find that the large-scale diffuse emission is more extended at the lower energies (<3 keV), implying that the optically thicker clouds responsible for scattering the hard continuum and Fe K$\alpha$ photons may be relatively more prevalent closer to the nucleus, as in the Milky Way. These clouds must not prevent soft ionizing X-rays from the AGN escaping to larger radii, in order to have photoionized ISM at larger radii. The diffuse emission is also significantly extended in the cross-cone direction, where the AGN emission would be mostly obscured by the torus in the standard AGN model, suggesting the possibility of a porous torus. However, in Paper III we show that the good correspondence between optical line, radio continuum and soft (<3 keV) X-ray features, is consistent with simulations of jet/molecular disk interaction; these simulations also suggest that this interaction may be responsible for the cross-cone extent.**

**Here we have shown (Section 4) that even within 100pc, there is extended and complex hard Fe K$\alpha$ emission.** If the extended emission from within 100pc to scales of ~5kpc is the result of the interaction of AGN photons with dense molecular clouds in the ISM, as argued in our previous work, its presence provides a measure of the nuclear activity time.

Based on the light travel time, we estimate that the AGN must have been active for at least $10^4$ yrs. Given the BH mass estimate for ESO 428-G014 of a few $10^7$ $M_\odot$ (Section 1.), **comparison with the models of Czerny et al. (2009) suggests that a long outburst of $10^4$ yr, as derived from the extent of the emission, may imply high accretion rates, at least Eddington (see Fig. 4 of Czerny et al. 2009).** In principle, comparisons of high sensitivity maps of the molecular gas and of the X-ray emission of ESO 428-G014 could establish the patterns of nuclear variability, as is being attempted for the Milky Way (e.g., Zhang et al. 2015; Churazov et al. 2017a).

**6. Conclusions**

High-resolution subpixel imaging of the hard continuum and Fe Kα emission of the CT AGN ESO 428-G014 observed with *Chandra* ACIS has uncovered differences in the morphology of the nuclear emission in different energy bands. In particular, a single nuclear point source is detected only in the 4-6 keV range. The 6-7 keV emission is characterized by two peaks of similar intensity, separated by ~36 pc in projection on the plane of the sky. Our analysis excludes astrometry problems, unknown systematics of the *Chandra* PSF, and statistical uncertainties as the cause of this result. Neither of the two Fe Kα knots is obviously coincident with the heavily obscured continuum source seen at 4 - 6 keV.

A large majority of the fluorescent Fe Kα arises >13 pc from the nucleus (see also Paper III). This result will need to be taken into account in models of the torus in ESO 0428-G014.

We discuss the implications of our results for four possible interpretations of the nuclear morphology:

(1) Given the bolometric luminosity and likely BH mass of ESO 428-G014, we may be imaging two clumps of the CT obscuring torus in the Fe Kα line. The observed Fe Kα emission in the central 100 pc of ESO 428-G014 is closer to the expected radii for the torus than in NGC 4945 (Marinucci et al. 2012, 2017).

(2) It is also possible that the Fe Kα knots are connected with the fluorescent emission from the dusty bicone (Asmus et al. 2016), as seen in the Circinus galaxy (Arevalo et al. 2014).

(3) The double knots could be the light echo of a relatively recent nuclear outburst, as seen in the Milky Way in the proximities of SgrA*.

(4) Although this is **a less likely scenario, for completeness we discuss the possibility of d**etecting the rare signature of a pair of merging nuclei, especially if a minor merger is in progress (Capelo et al. 2017). As a CT AGN, ESO 428-G014 is a likely candidate for a 'buried' pre-blow-out phase in the merger cycle (Hopkins et al., 2008).

Finally, considering the total extent of the hard continuum and Fe Kα emission (Papers I and II), we conclude **(based on Czerny et al 2009),** that the AGN in ESO 428-G014 **may have been** active at high accretion rates for at least $10^4$ yrs.

The ultimate sub-arcsecond spatial resolution of *Chandra* is essential for pursuing these studies, as is the high-count statistics that can only be obtained with long *Chandra* exposures. These results strengthen the case for a future large-collecting-area X-ray observatory that preserves, or exceeds, the angular resolution of *Chandra*, such as *Lynx*.

We retrieved data from the NASA-IPAC Extragalactic Database (NED) and the *Chandra* Data Archive. For the data analysis, we used the *CIAO* toolbox, *Sherpa*, and *DS9,* developed by the *Chandra* X-ray Center (CXC), and *LIRA* ([github.com/astrostat/LIRA](github.com/astrostat/LIRA)). This work was partially supported by the *Chandra* Guest Observer program grant GO5-16090X (PI: Fabbiano) and NASA contract NAS8- 03060 (CXC). The work of J.W. was supported by the National Key R&D Program of China (2016YFA0400702) and the National Science Foundation of China (11473021, 11522323). This work was performed in part at the Aspen Center for Physics, which is supported by National Science Foundation grant PHY-1607611.